\documentclass{aa}

\usepackage[varg]{txfonts}

\usepackage{lineno}
\usepackage{natbib}
\usepackage{amsmath}
\usepackage{graphicx}
\usepackage{xspace}
\usepackage{xcolor}
\usepackage[markup=underlined]{changes}

\usepackage{hyperref}
\hypersetup{
    colorlinks=true,
    linkcolor={red!50!black},
    citecolor={blue!70!black},
    urlcolor={blue!80!black}
}

\graphicspath{{./figs/}{./subs/}}

\usepackage[normalem]{ulem}
\usepackage{multirow}

\usepackage{etoolbox}
\newtoggle{thesis}

\begin{document}

\title{Discovery of a quasi-periodic oscillation non-harmonically related to the Type-C QPO in the hard intermedidate state of  MAXI~J1820$+$070}
\titlerunning{Type-B QPO spanning the hard-intermediate and soft-intermediate states}

\author{Pei~Jin\inst{1}\thanks{peijin@astro.rug.nl} \and
Mariano~M\'{e}ndez\inst{1}\thanks{mariano@astro.rug.nl} \and
Federico~Garc\'{\i}a\inst{2} \and
Diego~Altamirano\inst{3} \and
Ruican~Ma\inst{3}}

\institute{Kapteyn Astronomical Institute, University of Groningen, P.O.\ BOX 800, 9700 AV Groningen, The Netherlands
\and Instituto Argentino de Radioastronom\'{\i}a (CCT La Plata, CONICET; CICPBA; UNLP), C.C.5, (1894) Villa Elisa, Buenos Aires, Argentina
\and School of Physics and Astronomy, University of Southampton, Southampton, Hampshire SO17 1BJ, UK
}

\date{Received xxx; accepted xxx}

\abstract{
We present a detailed timing analysis of the transition from the hard-intermediate state (HIMS) to the soft-intermediate state (SIMS) in MAXI J1820+070 using NICER observations. This transition is marked by a sharp drop of the broadband noise across both the soft and hard X-ray bands, the disappearance of the Type-C quasi-periodic oscillation (QPO), the quenching of the steady, optically thick, compact jet, the appearance of a Type-B QPO, and the detection of discrete, optically thin, radio ejections. For the first time, we detect a QPO at $\sim$3.5–5.9 Hz in the 2–12 keV power density spectrum of MAXI J1820+070 roughly half a day before the transition, which appears to evolve smoothly into the Type-B QPO observed immediately after the transition. The location of this additional QPO component in the broadband rms vs. QPO frequency plot is consistent with that of the Type-B QPOs in GX 339$-$4 and GRO J1655$-$40, suggesting a possible connection between this additional QPO in the HIMS and the Type-B QPO in the SIMS. This result, together with recent findings in Swift J1727.8$-$1613, suggests that QPOs with these characteristics can emerge prior to the HIMS-to-SIMS transition and are not confined exclusively to the SIMS. If this additional QPO feature is the precursor of the Type-B QPO in the SIMS, its presence before the transition, whereas the bright discrete, optically thin, radio ejections appear at the transition, would imply that there may be no direct physical connection between the Type-B QPO and the discrete radio ejections. Our results further suggest a link between the disappearance of the Type-C QPO, the drop of the broadband noise, and the emergence of discrete radio ejections at the HIMS-to-SIMS transition. We speculate that the simultaneous presence of such a QPO, non-harmonically related to the Type-C QPO in the HIMS, could be compatible with a spine–sheath outflow structure.
}


\maketitle 



\section{Introduction}
\label{sec:introduction}

Black-hole X-ray binaries (BHXBs) are generally transient systems that remain in a quiescent state most of the time, and occasionally undergo bright X-ray outbursts lasting from several weeks to months~\citep[see, e.g.,][for review]{1996ARA&A..34..607T, 2006csxs.book..157M}. During these outbursts, BHXBs X-ray spectra typically consist of two main components: a thermally emitting component and a power-law (PL) tail~\citep[e.g.,][]{1997ApJ...479..926M}. The thermal component arises from the accretion disk~\citep{1973A&A....24..337S} and can be modeled as a multi-temperature blackbody with a characteristic temperature of about 0.3-2~keV~\citep{1984PASJ...36..741M}. In contrast, the PL component is attributed to Comptonization in a hot corona, where soft photons are up-scattered by high-energy electrons with temperatures reaching up to $\sim$100~keV~\citep[e.g.,][]{1980A&A....86..121S}.

BHXBs exhibit significant variability over a wide range of time scales, from milliseconds to hundreds of seconds~\citep[see reviews by ][and references therein]{2011BASI...39..409B, 2019NewAR..8501524I, 2025Galax..13..111Z}. The most striking timing features in these sources are the quasi-periodic oscillations (QPOs), which manifest themselves as narrow peaks in the Fourier power density spectrum (PDS). Low-frequency QPOs, with frequencies typically between 0.1 and 30 Hz, are categorized into three types—A, B, and C—based on their centroid frequency, amplitude, width, and the properties of the associated broadband noise~\citep{1999ApJ...526L..33W, 2002ApJ...564..962R, 2005ApJ...629..403C}. At the same time, close to state transitions the phenomenology can be more complex, with multiple variability components coexisting or evolving rapidly, which can make the classification of individual features less straightforward.

The evolution of a BHXB during an outburst can be conveniently described using the hardness–intensity diagram (HID), in which the source follows a counter-clockwise “q”-shaped track~\citep[see reviews by ][and references therein]{2001ApJS..132..377H, 2005Ap&SS.300..107H, 2010LNP...794...53B, 2011BASI...39..409B}. Based both on spectral and timing characteristics, four main states are identified along this track: the Low Hard State (LHS), Hard Intermediate State (HIMS), Soft Intermediate State (SIMS), and High Soft State (HSS). At the beginning of an outburst, the system enters the LHS, corresponding to the nearly right vertical branch seen in the HID~\citep[see, e.g., Figure 5 in][]{2011BASI...39..409B}. In this state, the emission is dominated by the PL component, and the PDS shows strong broadband noise, with a large total fractional rms amplitude of about 30-40\% in the 2–15~keV band, and Type-C QPOs~\citep{1997ApJ...479..926M, 2005Ap&SS.300..107H, 2011MNRAS.410..679M}.

As the outburst evolves, the source transitions toward softer states near the outburst peak—namely the intermediate states, which correspond to the upper horizontal branch of the HID. During this phase, the disk emission increases as the disk temperature rises. The HIMS exhibits timing properties that evolve continuously from the LHS, characterized by an increase of characteristic variability frequencies and a reduction of the broadband rms amplitude~\citep{2001ApJS..132..377H, 2005Ap&SS.300..107H, 2011MNRAS.410..679M}. 
When the source crosses the so-called jet line~\citep{2004MNRAS.355.1105F, 2009MNRAS.396.1370F}, discrete radio ejections are detected in the radio band, the Type-C QPO disappears, and the strength of the broadband noise decreases significantly. These changes indicate that the source has transitioned into the SIMS~\citep[e.g.,][]{2005Ap&SS.300..107H, 2011BASI...39..409B, 2011MNRAS.410..679M, 2012MNRAS.421..468M, 2019ApJ...883..198R, 2020ApJ...891L..29H}.
In the SIMS, the Type-B QPO dominates the X-ray PDS. 
This phenomenology suggested that the Type-B QPO was associated to the discrete radio ejections, and produced by precession of the jet~\citep{2016MNRAS.460.2796S, 2020A&A...640L..16K}.
However, very recently, \citet{2026A&A...706A.208J} detected a Type-B QPO in the HIMS and suggested that the discrete radio ejection is associated with the disappearance of the Type-C QPO, rather than the appearance of the Type-B QPO.
In the HSS, corresponding to the left vertical branch of the HID, the spectrum becomes dominated by thermal emission from the accretion disk, and the PL component weakens significantly or vanishes~\citep{1997ApJ...479..926M, 2005Ap&SS.300..107H, 2010LNP...794...53B, 2024MNRAS.530..929J}. Occasionally, weak Type-A QPOs are observed during the SIMS/HSS~\citep{2004A&A...426..587C, 2011MNRAS.418.2292M, 2023MNRAS.526.3944Z}.

MAXI~J1820$+$070 (ASASSN-18ey) is a high-inclination ($>63^\circ$) low-mass X-ray binary hosting a stellar-mass black hole of 5.7$-$8.5$\,M_\odot$~\citep{2020ApJ...893L..37T}. It was first identified in X-rays by MAXI on 2018 March 11, shortly after its optical counterpart was detected by ASAS-SN \citep{2018ATel11399....1K}. During its 2018 outburst, the source underwent an extended LHS, transitioned through intermediate states, and entered the HSS before returning to the LHS, with a transient discrete radio ejection occurring at the transition from the HIMS to the SIMS~\citep{2020ApJ...891L..29H, 2020NatAs...4..697B, 2021MNRAS.505.3393W}.

The transition from the HIMS to the SIMS in MAXI~J1820$+$070 took place at MJD $\sim$58305.67740~\citep{2020ApJ...891L..29H}, marked by a significant drop in the 0.3$-$12 keV broadband noise level and a switch from the Type-C QPO, which was present in the LHS and the HIMS, to the Type-B QPO. Contemporaneous radio monitoring of MAXI~J1820$+$070 revealed a dramatic jet evolution around the HIMS-to-SIMS transition, with the steady compact jet being rapidly quenched and the appearance of a bright radio flare associated with discrete relativistic ejections~\citep{2020NatAs...4..697B}.
Based on a study of the fractional-rms and phase-lag spectra of the QPOs before and after the transition, \citet{2023MNRAS.525..854M} suggested that a horizontally extended corona transitions into a vertically extended corona during this state change. In addition, the high-frequency covariance spectrum \citep{2021A&A...654A..14D} indicated that, near the transition, the disk inner radius moved to, or very close to, the innermost stable circular orbit~\citep[see also][]{2025MNRAS.538.1143L}.

In this paper, we present a detailed timing analysis of NICER observations around the transition from the HIMS to the SIMS reported by~\citet{2020ApJ...891L..29H}. The main difference between our and previous studies of the transition is that we fit the power and cross spectra in different energy bands following the procedure introduced by~\citet{2024MNRAS.527.9405M}. The paper is organized as follows: In Section~\ref{sec:OBSERVATION AND DATA REDUCTION} we describe the data reduction and analysis, in Section~\ref{sec:results} we show the results from the timing analysis, which we discuss in Section~\ref{sec:discussion}.

\section{Observations and data reduction}
\label{sec:OBSERVATION AND DATA REDUCTION}

The Neutron star Interior Composition Explorer~\citep[NICER,][]{2016SPIE.9905E..1HG} observed the outburst of MAXI~J1820$+$070 from 2018 March 6 (MJD 58183) to 2018 November 21 (MJD 58443). In this work, we focus on ObsID 1200120197, obtained on MJD 58305, which covers the transition from the HIMS to the SIMS~\citep{2020ApJ...891L..29H, 2023MNRAS.525..854M}. We use HEASOFT~v6.35 and CALDB~v20220331 to process the data, in particular we use {\tt nicerl2} to reprocess the data and generate a clean event file, and {\tt nicerl3-lc} to produce the light curve. We do not subtract the background count rate from the light curve, as the source is bright and the background contribution is relatively negligible.

We use GHATS v3.3.0\footnote{https://github.com/ghats-timing/ghats} to compute the Fast Fourier Transform for segments of 64~s or 16~s, with a Nyquist frequency of 500~Hz, in order to produce PDS in different energy bands and cross spectra (CS) of pairs of energy bands.
Depending on the scientific purpose, we select data within certain time intervals (see Fig.~\ref{fig:light_curve}) and average the PDS and CS from the relevant segments to obtain averaged PDS and CS.
We then rebin the averaged PDS and real and imaginary parts of the CS in frequency by a factor of $10^{1/100} \approx 1.023$ to increase the signal-to-noise ratio further, still maintaining a good frequency resolution. In Section~\ref{sec:Dynamical power spectrum}, we adopt a coarser rebinning factor of $10^{1/50} \approx 1.047$.
The PDS and the real and imaginary parts of the CS are normalized to fractional rms-squared units~\citep{1990A&A...230..103B}. The Poisson noise is estimated in the 200$-$500 Hz frequency range where no source variability is observed, and subtracted. The background count rate is not considered when we compute the rms normalization, since it is negligible compared to the source count rate.

We use Xspec~v.12.14.0~\citep{1996ASPC..101...17A} to fit the PDS and the real and imaginary parts of the CS.
In Section~\ref{sec:Power and cross spectra}, we simultaneously fit the $0.5-1.5$~keV and $2.0-12.0$~keV PDS, as well as the real and imaginary parts of the corresponding CS, using a combination of Lorentzian functions, as proposed by~\cite{2024MNRAS.527.9405M}.  

The method introduced by~\citet{2024MNRAS.527.9405M} relies on four assumptions:
(i) the PDS of these sources can be described by a linear combination of Lorentzian functions;
(ii) the centroid frequency and full width at half maximum (FWHM) of each Lorentzian are identical in all energy bands;
(iii) each Lorentzian is perfectly coherent between any two energy bands;
(iv) any two Lorentzians are incoherent with one another, at least when they overlap in frequency.
Assumption (i) is well motivated, since it is standard practice to model the PDS of BHXBs with a linear combination of Lorentzian functions~\citep[e.g.;][]{2000MNRAS.318..361N, 2002ApJ...572..392B}.
Assumption (ii) is supported by the fact that in the majority of sources, the centroid frequency of the QPO is consistent across all energy bands up to 100 keV~(e.g., MAXI~J1820$+$070, \citealt{2021NatAs...5...94M,2023ApJ...948..116M}; Swift~J1727.8$-$1613, \citealt{2024ApJ...970L..33Y}).
In the few cases in which there are reports that the QPO frequency changes with energy~\citep[e.g.;][]{2010ApJ...710..836Q, 2013MNRAS.428.1704L, 2013MNRAS.433..412L, 2018MNRAS.474.1214Y}, it has been shown~\citep{2024MNRAS.527.9405M, 2026A&A...706A.208J} that for 4 or more energy bands, a fit of the QPO feature with two Lorentzians whose frequencies and FWHM are the same in all energy bands is statistically better and requires less parameters than the alternative with one Lorentzian with centroid frequency and FWHM that depend on energy~\citep[see Sec. 3.3 of][for details]{2024MNRAS.527.9405M}.
Assumption (iii) is consistent with the fact that the coherence function of the kilo-Hertz QPOs in neutron-star systems, which appear in regions of the PDS where no other component contributes to the variability, show unit coherence~\citep[e.g.;][]{1997ApJ...483L.115V, 1999ApJ...514L..31K, 2013MNRAS.433.3453D, 2018ApJ...860..167T}.
The same happens with strong Type-C QPOs in BHXBs. When the Type-C QPO dominates the PDS, the coherence function near its peak is close to unity. This is seen, for example, in XTE~1550$-$564~\citep{2000ApJ...531L..45C, 2017MNRAS.469.2011R} and Swift~J1727.8$-$1613~\citep{2025A&A...699A...9J}. In these sources, coherence falls below one in the wings of the profile, where other components become comparable in strength. As explained in~\citet[][see also assumption iv below]{1997ApJ...474L..43V}, overlap between multiple mutually incoherent components lowers the measured coherence. By contrast, coherence remains close to unity at the centroid frequency, where the Type-C QPO is strongest.
Assumption (iv) predicts a drop in the coherence function when multiple components that are incoherent with one another overlap in frequency~\citep{1997ApJ...474L..43V}, as reported in several sources (see below).

We emphasize that the method of~\citet{2024MNRAS.527.9405M} is a model, and as such it depends upon the validity of the four assumptions described above. The fact that the data might be described by other existing models~\citep[e.g.;][]{2013MNRAS.434.1476I, 2016MNRAS.458.3655V, 2022MNRAS.515.1914Z}, or by other models to be proposed in the future, does not invalidate this model or its assumptions. The validity of these assumptions can be judged by whether the model fits the data and makes predictions that can be verified. We note that, when applied to the power and cross spectra, this model correctly predicts the phase-lag spectrum and the coherence function~\citep[e.g.;][]{2024MNRAS.527.9405M, 2025A&A...696A.128B, 2025A&A...699A...9J, 2026A&A...706A.208J, 2025A&A...696A.237F, 2025arXiv250507938B, 2025ApJ...990...43R}, and has been able to reproduce and explain narrow drops in the coherence function in several sources~\citep{2024MNRAS.527.9405M, 2024A&A...687A.284K, 2025A&A...696A.128B, 2025A&A...696A.237F, 2025arXiv250507938B, 2025ApJ...990...43R}, as well as distinctive features in the lag spectrum~\citep{2026A&A...706A.208J}.

During the fit, the phase lag of each Lorentzian function is assumed to be constant with frequency, and we link the centroid frequency and FWHM for each Lorentzian function across the different spectra (assumption ii). Each Lorentzian component therefore has six free parameters: the centroid frequency $\nu_0$, FWHM, which, as we explained, are assumed to be the same in all four spectra, the phase lag between the two energy bands, and three normalization factors for the $0.5-1.5$~keV, and $2.0-12.0$~keV PDS, and the CS.
We confirmed that the fitted normalizations are always consistent within errors with the relations expected from the mathematical formalism~\citep[equation 7,][]{2024MNRAS.527.9405M}. We consider a Lorentzian component to be significant only if it exceeds a 3$\sigma$ significance level in at least one of the PDS or the CS.

To generate fractional-rms and phase-lag spectra of the QPOs, we divide the full energy range into 8 bands following~\citet{2023MNRAS.525..854M}. The separate energy bands are 0.5$-$0.75 keV, 0.75$-$1.0 keV, 1.0$-$1.5 keV, 1.5$-$2.5 keV, 2.5$-$4.0 keV, 4.0$-$5.0 keV, 5.0$-$6.5 keV, and 6.5$-$12.0 keV.
For Intervals \#3, \#4, and \#5 (Fig.~\ref{fig:light_curve}), we instead use four broader energy bands, 0.5$-$1.5 keV, 1.5$-$3.5 keV, 3.5$-$6.0 keV, and 6.0$-$12.0 keV, to improve the signal-to-noise ratio.
We use the 0.5$-$12.0 keV band as reference and each of the other bands as subject to produce the corresponding CS. 
We correct for the partial correlation introduced by the fact that the subject band is part of the reference band following the procedure described in \citet[][see also \citealt{2024MNRAS.527.7136B}]{2019MNRAS.489.3927I}.
In this manner, we obtain eight PDS and eight CS for most observations, but four PDS and four CS for Intervals~\#3, \#4 and~\#5 shown in Fig.~\ref{fig:light_curve}.
We fit jointly each of the eight (four) groups of PDS and real and imaginary parts of the CS using a constant phase-lag model~\citep{2024MNRAS.527.9405M}. As in the broadband analysis, we also link the centroid frequency and FWHM for each Lorentzian function across the different spectra in the fit.
We subsequently calculate the 1-$\sigma$ confidence range of the parameters using the Markov chain Monte-Carlo algorithm (MCMC). 
The Goodman-Weare algorithm is applied for a total of 200000 samples and a burn-in phase long enough to ensure that the chain reaches a steady state.

\section{Results}
\label{sec:results}

\subsection{Transition from the HIMS to the SIMS}
\label{sec:Transition from the HIMS to the SIMS}

\begin{figure}
	\includegraphics[width=\columnwidth]{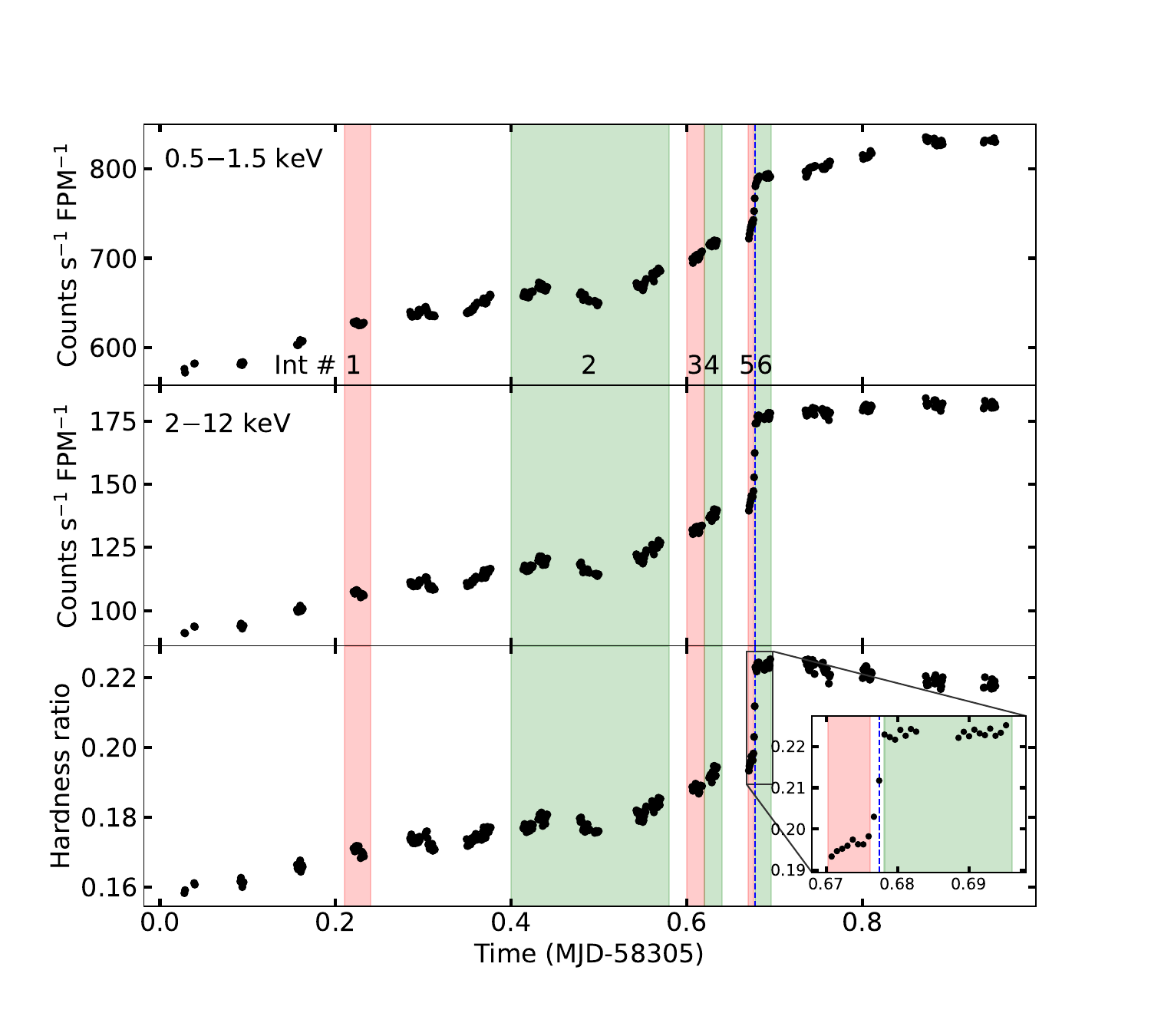}
    \caption{X-ray Light curves and hardness ratio of MAXI~J1820$+$070 for ObsID 1200120197 on MJD 58305. Upper panel: the 0.5$-$1.5~keV light curve. Middle panel: the 2.0$-$12.0~keV light curve. Bottom panel: the hardness ratio between the 2.0$-$12.0 keV and 0.5$-$1.5~keV light curves. The blue dotted line marks the transition from the HIMS to the SIMS~\citep{2020ApJ...891L..29H}. The shaded regions indicate the intervals for which we apply the joint-fitting method to model the PDS and CS. Each point is 64 seconds.}
    \label{fig:light_curve}
\end{figure}

In Fig.~\ref{fig:light_curve}, we present the X-ray light curves in the 0.5$-$1.5~keV (upper panel) and 2.0$-$12.0~keV (middle panel) bands, along with the corresponding hardness ratio (2.0$-$12.0~keV / 0.5$-$1.5~keV; bottom panel) of MAXI~J1820$+$070 for ObsID 1200120197 on MJD 58305.
The light curves are normalized to counts per Focal Plane Module (FPM) with each point being 64-seconds long.
During the observation, the 0.5$-$1.5~keV light curve increases from $\sim$572 to $\sim$835 counts~s$^{-1}$~FPM$^{-1}$, while the 2.0$-$12.0~keV light curve increases from $\sim$91 to $\sim$185~counts~s$^{-1}$~FPM$^{-1}$. Consequently, the hardness ratio increases from $\sim$0.16 to $\sim$0.23.
At MJD $\sim$58305.67740, indicated by the blue dotted line, the source underwent a transition from the HIMS to the SIMS~\citep{2020ApJ...891L..29H}, characterized by the following phenomena:
 (i) a switch from the Type-C to the Type-B QPO~\citep{2020ApJ...891L..29H, 2023MNRAS.525..854M};
 (ii) a drop in the broadband noise level in the power spectrum~\citep{2020ApJ...891L..29H};
 (iii) a bright radio ejection event~\citep{2020ApJ...891L..29H, 2020NatAs...4..697B, 2021MNRAS.505.3393W}.

\subsection{Dynamical power spectrum}
\label{sec:Dynamical power spectrum}

\begin{figure*}
	\includegraphics[width=0.50\textwidth]{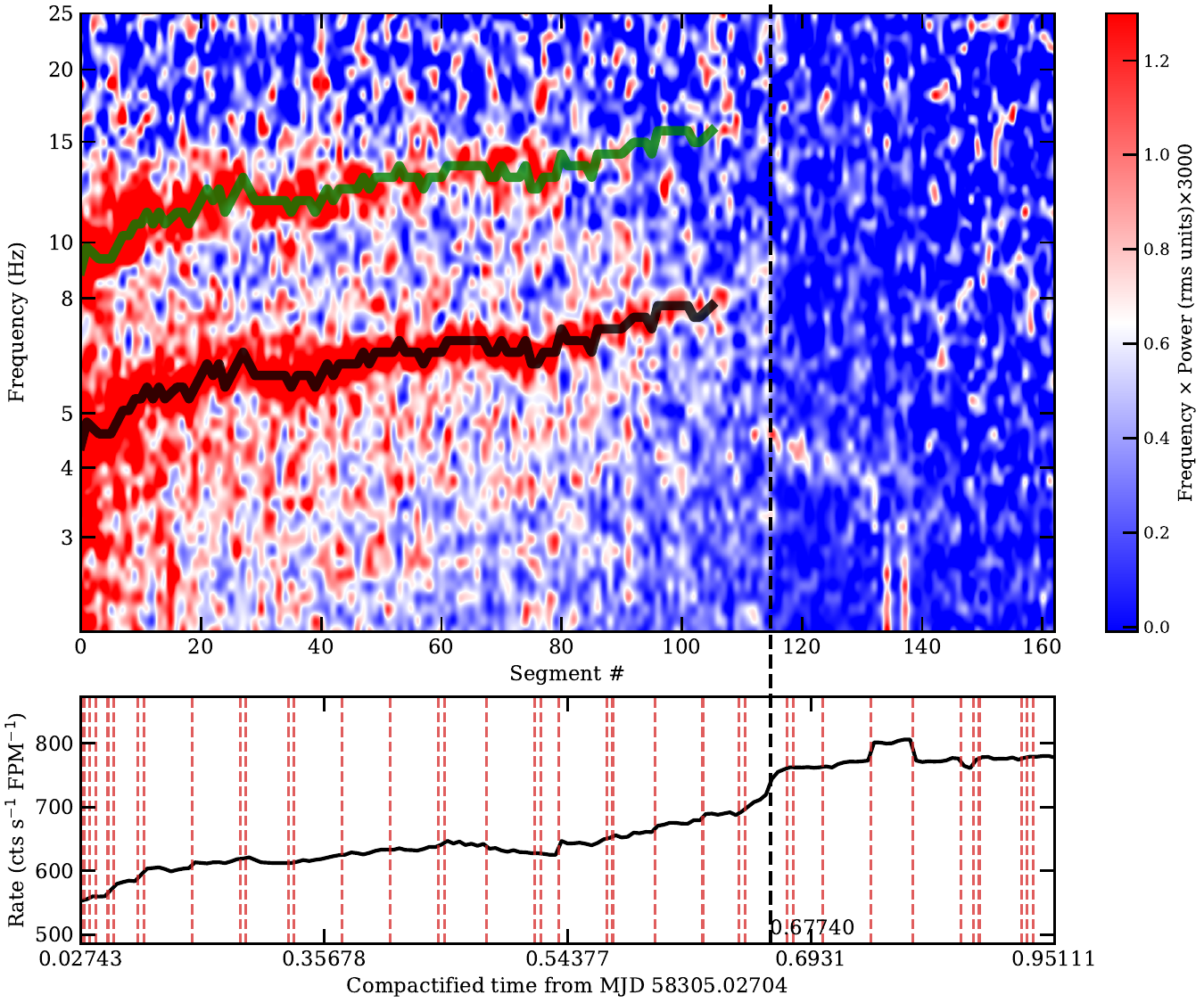}
    \includegraphics[width=0.50\textwidth]{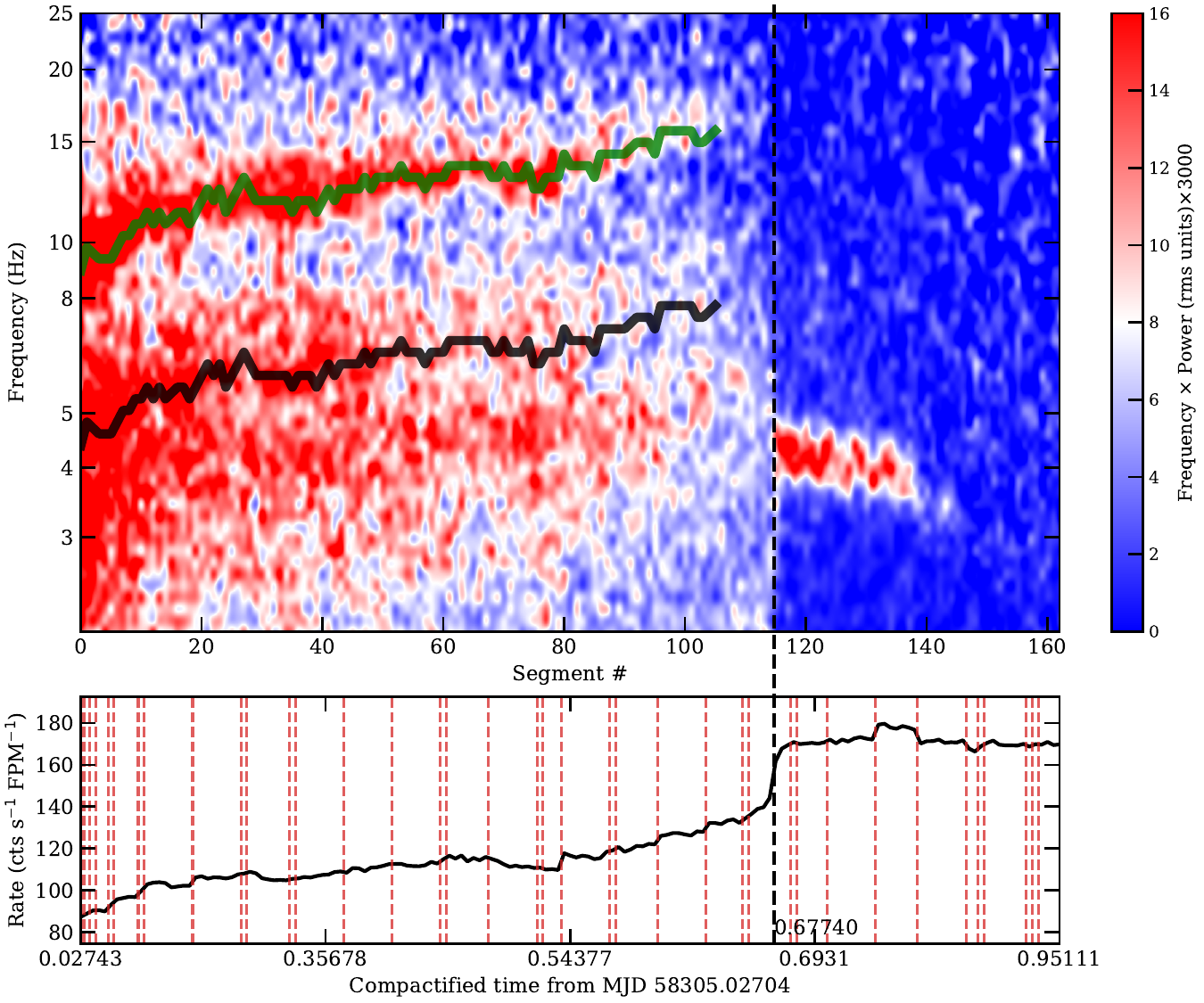}
    \caption{Dynamical power spectra (top panels) and light curves (bottom panels) of MAXI~J1820$+$070 for ObsID 1200120197: 0.5$-$1.5 keV (left) and 2.0$-$12.0 keV (right). The segments in the X-axis of the dynamical PDS have a duration of $\sim$128 seconds. Compactified time in the bottom panels is the true elapsed time since MJD 58305.02704, with the gaps in the data removed from the plot. We do not show the time gaps between observations to produce a more compact plot; the time gaps can be seen in Fig.~\ref{fig:light_curve}. The black and green lines in both panels indicate the trace of the frequency of the Type-C QPO and its harmonic, respectively. The black vertical line marks the time of the HIMS-to-SIMS transition. The red vertical lines mark the presence of time gaps.}
    \label{fig:dynamical_pds}
\end{figure*}

In Fig.~\ref{fig:dynamical_pds}, we present the dynamical power spectra in the 0.5$-$1.5 keV (left panel) and 2.0$-$12.0 keV (right panel) bands of ObsID 1200120197,
where the colour scale indicates the power as a function of time and frequency.
Two prominent and narrow features, indicated by the black and green lines, are clearly visible in both panels, which we identify as the Type-C QPO and its second harmonic, by comparison with the 0.3$-$12.0 keV dynamical power spectrum reported by \citet{2020ApJ...891L..29H}. Superimposed on these narrow features, a broad distribution of power extending over a wide frequency range, up to at least 15 Hz, is present, which we identify as the broadband noise.
At MJD$\sim$58305.67740 (a compactified time of 0.67740), this broadband component drops significantly in both dynamical power spectra, especially in the 2$-$12 keV band. Before this time, the Type-C QPO and its second harmonic are present in both spectra, with the Type-C QPO frequency increasing from  $\sim$4.5 Hz to $\sim$8.0 Hz. At the same time the frequency of the second harmonic increases from $\sim$9 Hz to $\sim$17 Hz. Just before that time, the Type-C QPO and its harmonic disappear in both spectra, with $\sim$95\% upper limits of 0.5\% and 0.4\% in the 0.5$-$12 keV band (Interval~\#5 in Table~\ref{tab:rep_obs}), respectively. Immediately after that time, both dynamical power spectra exhibit a pattern typical of the SIMS, characterized by a very weak broadband noise and a Type-B QPO with a frequency of $\sim$3.3$-$4.2 Hz.
This transition is also evident in the 0.3$-$12.0 keV dynamical power spectrum~\citep[Figure 2 of][]{2020ApJ...891L..29H}, marking that the source transitions from the HIMS to the SIMS.
We further note that, before the transition, the 2.0$-$12.0 keV dynamical power spectrum (Fig.~\ref{fig:dynamical_pds}, right panel) shows a variability component with a characteristic frequency of $\sim$3.5$-$5.9 Hz, appearing as a peak feature; hereafter we will refer to this feature as the additional QPO component (see Section~\ref{sec:Power and cross spectra}).

\subsection{Power and cross spectra}
\label{sec:Power and cross spectra}

\begin{figure*}
	\includegraphics[width=0.45\textwidth]{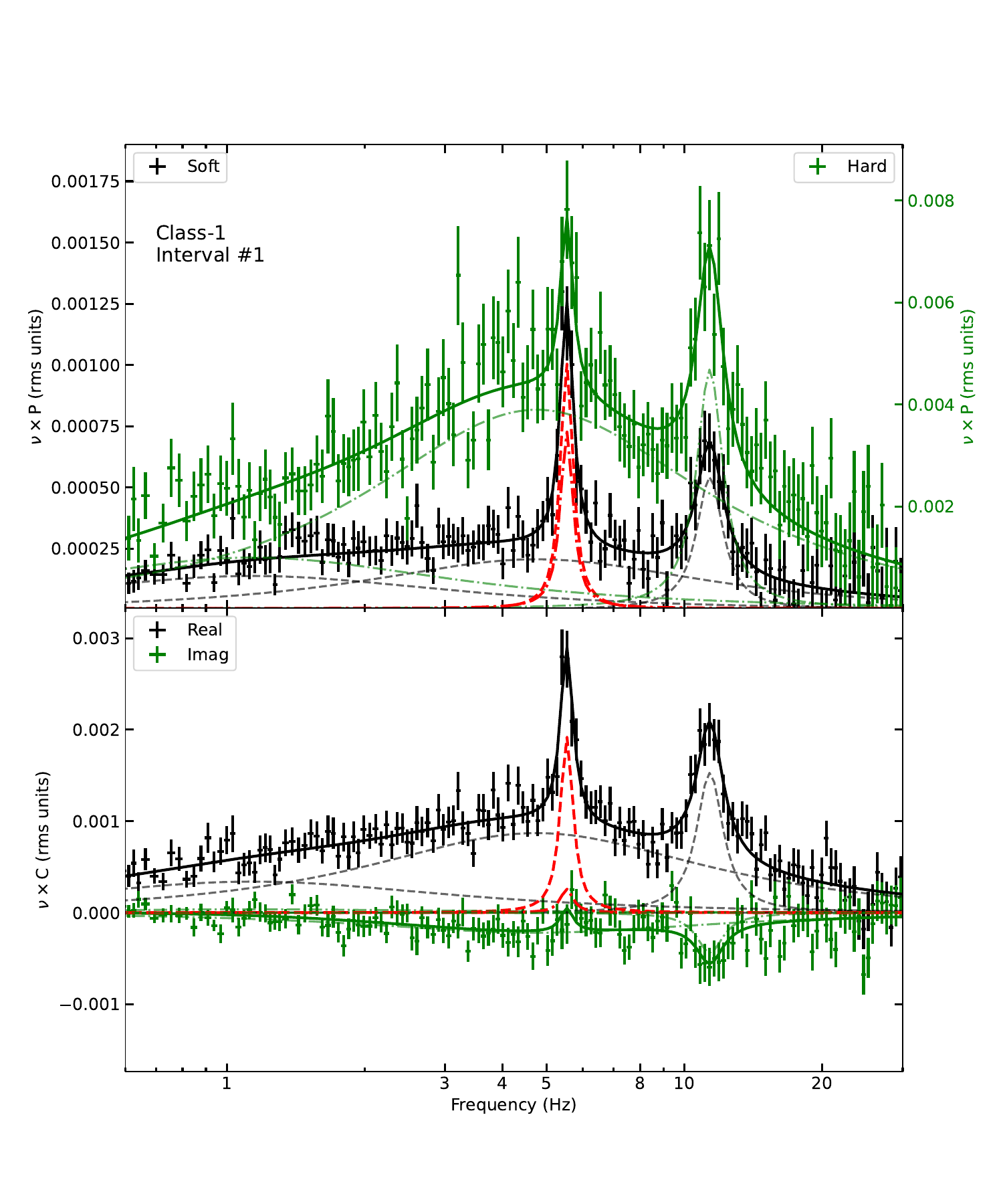}%
    \includegraphics[width=0.45\textwidth]{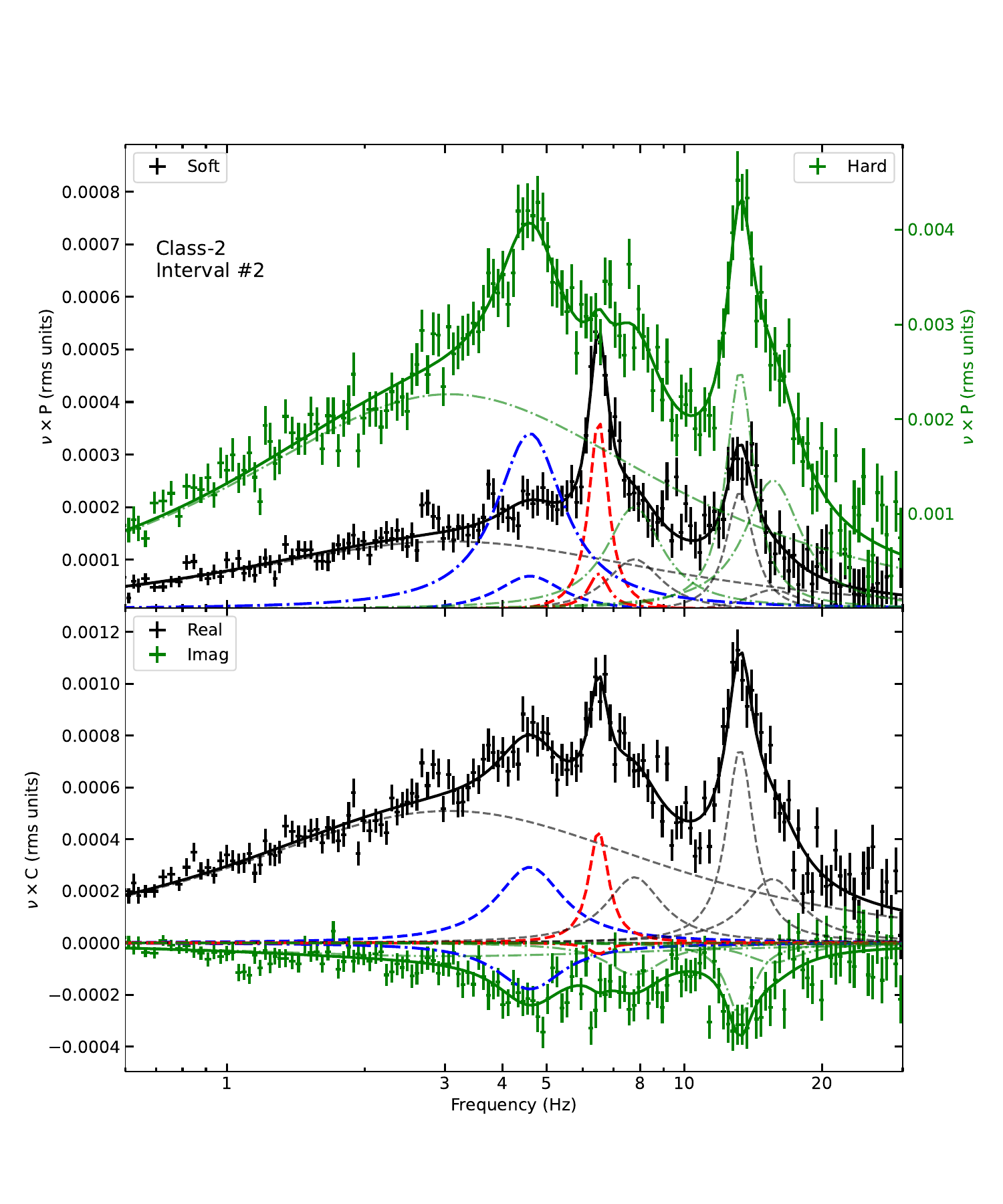}\vspace{-5mm}
    \includegraphics[width=0.45\textwidth]{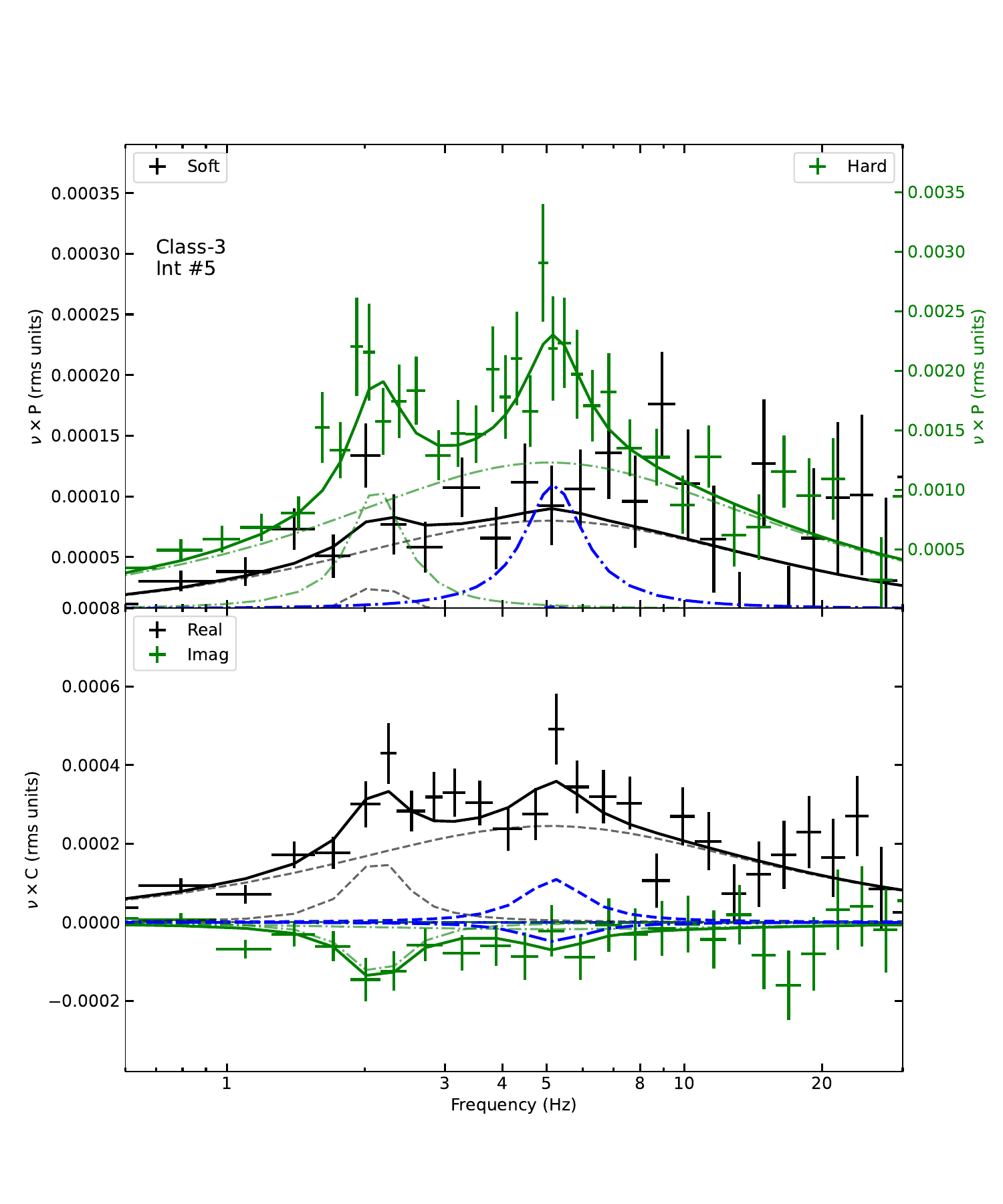}%
    \includegraphics[width=0.45\textwidth]{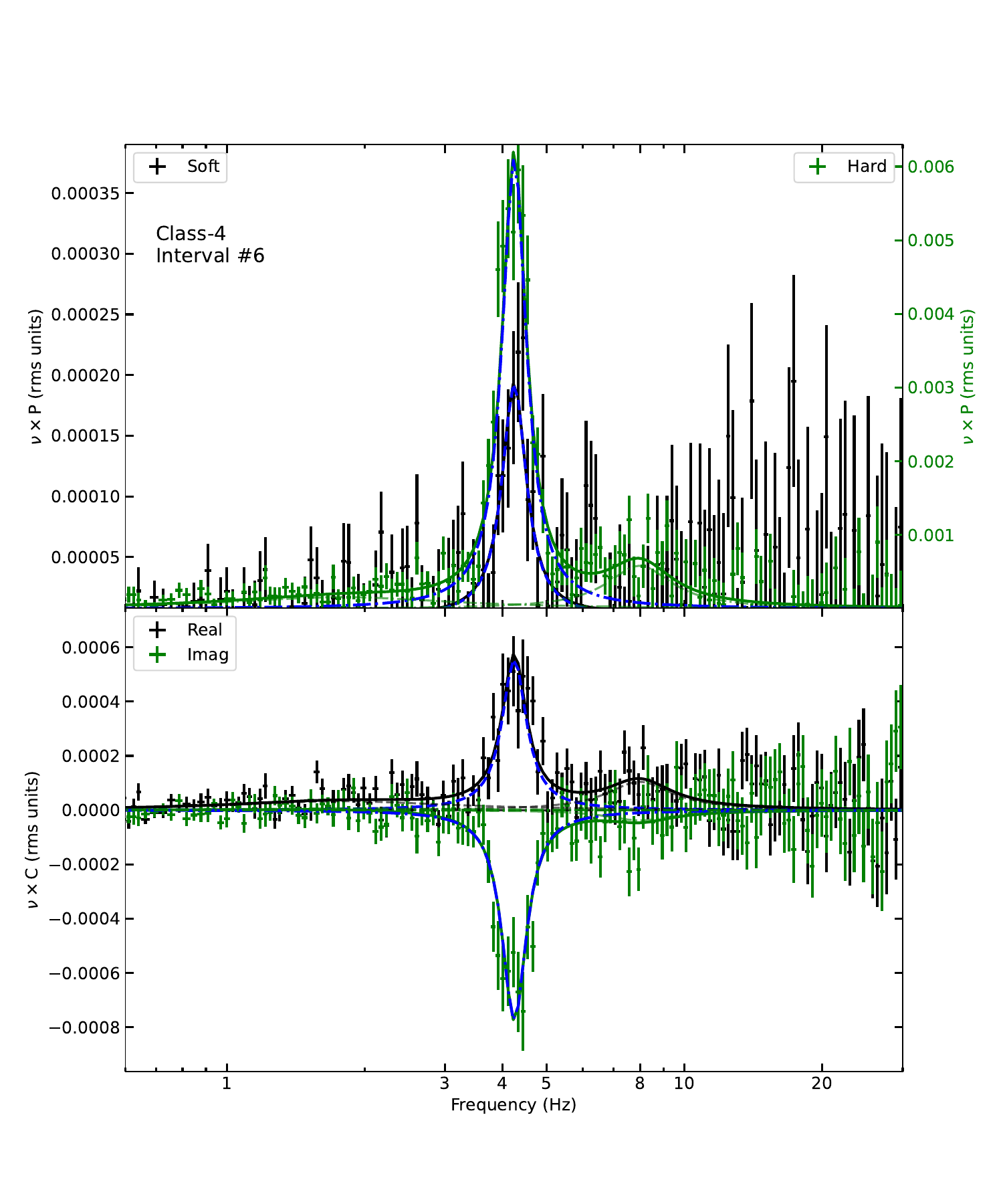}
    \caption{PDS and real and imaginary parts of the CS for four PDS classes near the transition from the HIMS to the SIMS in MAXI~J1820$+$070: Class-1 in the upper-left panel (Interval~\#1), Class-2 in the upper-right panel (Interval~\#2), Class-3 in the lower-left panel (Interval~\#5), and Class-4 in the lower-right panel (Interval~\#6).
    The soft band is 0.5$-$1.5 keV and the hard band is 2.0$-$12.0 keV. To improve visibility, we scale the soft-band PDS by the multiplicative constants indicated in the legends. We plot the contribution of each Lorentzian to the PDS and CS and highlight the Type-C QPO in red and the additional QPO component in the HIMS and the Type-B QPO in blue. In the lower-left panel, we rebin the plot by an extra factor of 5.}
    \label{fig:pds_cs}
\end{figure*}

\begin{table*}
    \centering
    \caption{Six representative intervals near the transition from the HIMS to the SIMS in MAXI~J1820$+$070, indicated by the shaded regions in Fig.~\ref{fig:light_curve}. }
    \label{tab:rep_obs}
    \begin{tabular}{lcccccc}
        \hline
        Interval& \#1 & \#2 & \#3 & \#4 & \#5 & \#6 \\
        \hline
        Start time (MJD-58305) & 0.22011 & 0.41316 & 0.60622 & 0.62521 & 0.67056 & 0.67814 \\
        Number of segments $\times$duration$^{*}$& 17$\times$65.536 s & 88$\times$65.536 s & 66$\times$16.384 s & 54$\times$16.384 s & 32$\times$16.384 s & 17$\times$65.536 s \\
        PDS Class & Class-1 & Class-2 & Class-2 & Class-2 & Class-3 & Class-4 \\
        \hline
        \multicolumn{7}{c}{Type-C QPO} \\
        $\nu_0$ (Hz) & $5.53_{-0.01}^{+0.02}$& $6.48_{-0.03}^{+0.04}$& $7.55_{-0.05}^{+0.09}$
& $7.9_{-0.2}^{+0.1}$& -- & -- \\
        FWHM (Hz)& $0.41_{-0.02}^{+0.11}$& $0.75 \pm 0.07$& $0.7_{-0.1}^{+0.2}$& $0.8_{-0.4}^{+0.2}$& -- & -- \\
        Phase lag (rad) & $0.13_{-0.10}^{+0.08}$& $-0.1 \pm 0.1$ & $0.9_{-0.5}^{+0.4}$& $1.6_{-0.7}^{+0.3}$& -- & -- \\
        rms 0.5$-$1.5 keV (\%)& $1.09_{-0.06}^{+0.09}$& $0.81_{-0.05}^{+0.07}$& $0.66_{-0.06}^{+0.05}$& $0.55_{-0.09}^{+0.02}$& $\leq 0.5^{\dagger}$& $\leq 0.2^{\dagger}$\\
        rms 2.0$-$12.0 keV (\%)& $2.0 \pm 0.2$
& $0.8_{-0.2}^{+0.3}$& $1.1 \pm 0.4$& $\leq 0.9^{\dagger}$& $\leq 1.0^{\dagger}$& $\leq 0.7^{\dagger}$\\
        covariance rms$^{**}$ (\%)& $1.5 \pm 0.1$& $0.9 \pm 0.1$& $0.5_{-0.2}^{+0.1}$& 
$0.56 \pm 0.02$& $\leq0.6^{\dagger}$& $\leq 0.7^{\dagger}$\\
        rms 0.5$-$12.0 keV (\%) & $1.20 \pm 0.05$ & $0.78 \pm 0.03$ & $0.60 \pm 0.05$ & $0.44 \pm 0.08$ & $\leq0.5^{\dagger}$& $\leq0.3^{\dagger}$\\
        \hline
        \multicolumn{7}{c}{Second harmonic of the Type-C QPO} \\
        $\nu_0$ (Hz) & $11.33_{-0.04}^{+0.08}$& $13.18_{-0.05}^{+0.06}$& $16.3_{-0.1}^{+0.2}$
& $17.0 \pm 0.6$
& -- & -- \\
        FWHM (Hz)&  $1.8 \pm 0.2$& $2.2_{-0.3}^{+0.4}$& $3.9_{-0.1}^{+0.6}$& $4.8_{-0.9}^{+1.2}$& -- & -- \\
        Phase lag (rad) &   $-0.28_{-0.05}^{+0.08}$& $-0.36_{-0.04}^{+0.03}$& $-0.17_{-0.14}^{+0.07}$& $-0.8 \pm 0.2$& -- & -- \\
        rms 0.5$-$1.5 keV (\%)& $1.16_{-0.08}^{+0.04}$& $0.77_{-0.07}^{+0.09}$& $0.49_{-0.13}^{+0.05}$& $0.8 \pm 0.1$& $\leq 0.5^{\dagger}$& $\leq 0.6^{\dagger}$\\
        rms 2.0$-$12.0 keV (\%)& $3.4 \pm 0.2$& $2.5 \pm 0.3$& $3.05_{-0.07}^{+0.22}$& $1.9 \pm 0.1$& $\leq 1.6^{\dagger}$& $\leq 0.5^{\dagger}$\\     
        covariance rms (\%)& $1.99_{-0.12}^{+0.04}$& $1.4 \pm 0.1$& $1.4 \pm 0.1$& $1.1 \pm 0.1$& $\leq1.0^{\dagger}$& $\leq 0.7^{\dagger}$\\
        rms 0.5$-$12.0 keV (\%)&  $1.59 \pm 0.04$& $1.12 \pm 0.02$ & $1.04 \pm 0.05$ & $0.95 \pm 0.06$ & $\leq0.4^{\dagger}$& $\leq0.5^{\dagger}$\\
        \hline
        \multicolumn{7}{c}{The additional QPO component in the HIMS or the Type-B QPO in the SIMS} \\
        $\nu_0$ (Hz) & -- & $4.50 \pm 0.03$& $5.1 \pm 0.2$ & $5.9 \pm 0.1$& $5.1_{-0.1}^{+0.3}$& $4.24 \pm 0.02$\\
        FWHM (Hz)& -- & $1.9 \pm 0.02$& $2.8_{-0.3}^{+0.8}$& $2.0_{-0.6}^{+0.3}$& $1.9 \pm 0.4$& $0.62_{-0.03}^{+0.02}$\\
        Phase lag (rad) & -- & $-0.55_{-0.04}^{+0.03}$& 
$-0.46_{-0.08}^{+0.18}$& $-0.80_{-0.2}^{+0.1}$& $-0.4_{-0.4}^{+0.5}$& $-0.95_{-0.09}^{+0.05}$\\
        rms 0.5$-$1.5 keV (\%)& $\leq 0.5^{\dagger}$& $0.65_{-0.06}^{+0.07}$& $0.65_{-0.06}^{+0.21}$& $0.4 \pm 0.2$& $0.23_{-0.03}^{+0.37}$& $0.66_{-0.03}^{+0.05}$\\
        rms 2.0$-$12.0 keV (\%)& $\leq 1.8^{\dagger}$& $3.4 \pm 0.02$& $3.7_{-0.4}^{+0.6}$& $2.3 \pm 0.3$& $2.4_{-0.3}^{+0.4}$& $3.7 \pm 0.1$\\
        covariance rms (\%)& $\leq 1.0^{\dagger}$& $1.45 \pm 0.07$& $1.7 \pm 0.2$& $1.1_{-0.2}^{+0.1}$& $0.8_{-0.2}^{+0.1}$& $1.47_{-0.08}^{+0.02}$\\
        rms 0.5$-$12.0 keV (\%)& $\leq0.9^{\dagger}$& $1.09 \pm 0.03$ & $1.21 \pm 0.06$ & $0.75 \pm 0.08$ & $0.66 \pm 0.12$& $1.07 \pm 0.03$ \\
        \hline
        \multicolumn{7}{p{0.95\linewidth}}{$^{*}$ For Intervals~\#1, \#2 and \#6, we compute the FFTs using segments of $\sim$64 s, whereas for Intervals~\#3, \#4, and \#5 we use shorter segments of $\sim$16 s.} \\
        \multicolumn{7}{p{0.95\linewidth}}{$^{**}$ the covariance rms is the magnitude of the QPO in the cross spectra of the 0.5$-$1.5 keV and 2.0$-$12.0 keV bands.} \\
        \multicolumn{7}{p{0.95\linewidth}}{$^{\dagger}$ The corresponding QPO is not detected and we give $\sim$95\% confidence upper limit of the fractional rms in each band and the covariance rms.} \\
    \end{tabular}
\end{table*}

We fit simultaneously the 0.5$-$1.5 keV and 2.0$-$12.0 keV PDS, as well as the real and imaginary parts of the corresponding CS, using the multi-Lorentzian method of~\citet{2024MNRAS.527.9405M}. The initial set of parameters is described in Section~\ref{sec:OBSERVATION AND DATA REDUCTION}.
Based on their properties, we categorize the PDS near the transition into four classes. To illustrate these classes, we select six representative intervals from the data shown in Fig.~\ref{fig:light_curve}. These intervals, indicated by the shaded regions in Fig.~\ref{fig:light_curve} and listed in Table~\ref{tab:rep_obs}, are used to present the corresponding PDS and CS.
In Fig.~\ref{fig:pds_cs}, we present the 0.5$-$1.5 keV (black) and 2.0$-$12.0 keV (green) PDS (upper panels), together with the real (black) and imaginary (green) parts of the CS (bottom panels) of the 2.0$-$12.0 keV data with respect to 0.5$-$1.5 keV data, for the four PDS classes. We plot the contribution of each Lorentzian to the PDS and CS in Fig.~\ref{fig:pds_cs}. We highlight the Type-C QPO in red and the additional QPO component in blue. In Table~\ref{tab:rep_obs}, we record the parameters of the QPOs.

\textit{(1) Class-1, before MJD 58305.25:}
In Interval~\#1 (Fig.~\ref{fig:pds_cs}, upper-left panel), the Type-C QPO at $5.53_{-0.01}^{+0.02}$ Hz and its harmonic at $11.33_{-0.04}^{+0.08}$ Hz dominate both the PDS and CS. The phase lag of the Type-C QPO is $0.13_{-0.10}^{+0.08}$ rad, while that of its harmonic is $-0.28_{-0.05}^{+0.08}$ rad. No other QPO peak is detected in this observation.

\textit{(2) Class-2, MJD 58305.25$-$58305.65:}
In Interval~\#2 (Fig.~\ref{fig:pds_cs}, upper-right panel), the 0.5$-$1.5 keV PDS shows a prominent Type-C QPO at $6.48_{-0.03}^{+0.04}$ Hz and its harmonic at $13.18_{-0.05}^{+0.06}$ Hz. However, in the 2.0$-$12.0 keV PDS, a narrow (FWHM$= 1.9 \pm 0.02$) QPO at $4.50 \pm 0.03$ Hz is more significant than the Type-C QPO. 
This additional QPO component appears in the high-energy PDS from MJD 58305.25 onward, as the strength of the Type-C QPO and that of its second harmonic decrease; before that, we do not detect this feature in either the PDS or the CS, with a $\sim$95\% confidence upper limit on its fractional rms of $\leq 1.8$\% in the 2.0$-$12.0~keV band for Interval \#1, immediately before the additional QPO component is detected. In the real part of the CS of Interval~\#2, three peaks are present, corresponding to the additional QPO component, the Type-C QPO and the harmonic of the Type-C QPO. The phase lag of the Type-C QPO is $-0.1 \pm 0.1$ rad, the phase lag of its harmonic is $-0.36_{-0.04}^{+0.03}$ rad, and that of the additional QPO component is $-0.55_{-0.04}^{+0.03}$ rad.
In Fig.~\ref{fig:pds_cs_obs3_4}, we present the results from Intervals~\#3 (upper panel) and \#4 (lower panel). Both Intervals~\#3 and \#4 are classified as Class-2 and exhibit a PDS and real and imaginary parts of the CS similar to those in Interval~\#2, but with higher characteristic frequencies of the variability components.

\begin{figure}
	\includegraphics[width=0.9\columnwidth]{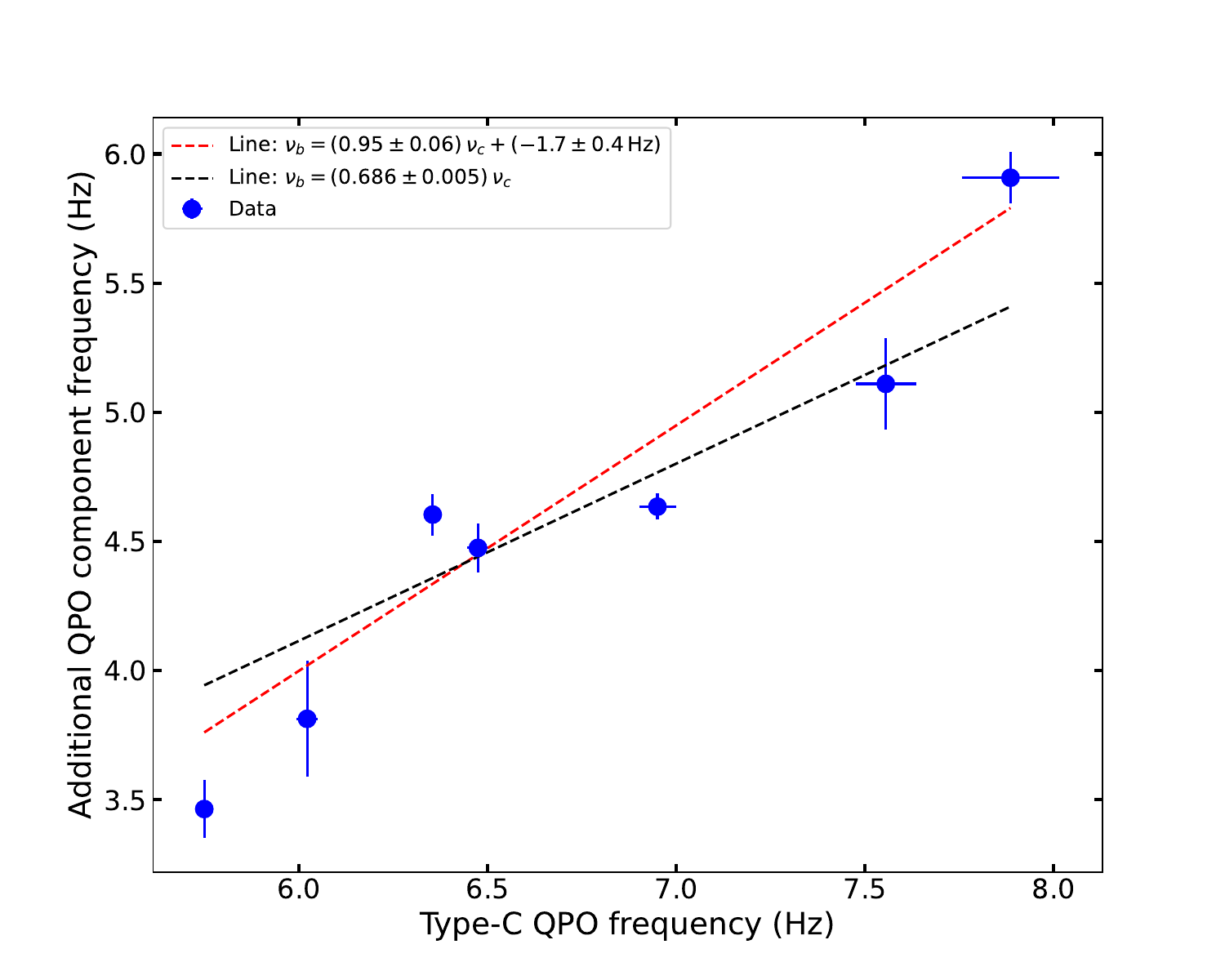}
    \caption{Frequency of the additional QPO component versus that of the Type-C QPO in MAXI~J1820$+$070 when both features appear simultaneously. The red line is the best fitting linear model to the data. The black line shows the fit with a fixed intercept of zero Hz.}
    \label{fig:frequency_ratio}
\end{figure}

In Fig.~\ref{fig:frequency_ratio}, we plot the frequency of the additional QPO component against that of the Type-C QPO for the Class-2 observations in which both features are detected simultaneously.
We fitted the data with a linear model. The best fit gives a slope of $0.95 \pm 0.06$, which differs from 0.5 by $\sim 8 \sigma$, and an intercept of $-1.7 \pm 0.4$ Hz, which differs from 0 by $\sim 4 \sigma$.
This indicates that the frequency ratio between the additional QPO component and the Type-C QPO is not constant. 
We also show in Fig.~\ref{fig:frequency_ratio} the best-fit model assuming a constant frequency ratio, with averaged frequency ratio of $0.686 \pm 0.005$.
These results indicate that the additional QPO component is unlikely to be directly linked to the Type-C QPO through a simple harmonic or scaling relation.

\textit{(3) Class-3, MJD 58305.67$-$58305.6774:}
In Interval~\#5 (Fig.~\ref{fig:pds_cs}, lower-left panel), taken just before the transition, the Type-C QPO becomes undetectable, with a $\sim$95\% upper limit of the rms amplitude in the 0.5$-$12 keV band of $\leq0.5$\%, and an upper limit of the QPO second harmonic of $\leq0.4$\%. The additional QPO component at $5.1_{-0.1}^{+0.3}$ remains significantly present in the high-energy PDS, with a significance of 3.8$\sigma$.
The phase lag of the additional QPO component is $-0.4_{-0.4}^{+0.5}$ rad in this observation.
The fractional-rms spectrum of this additional QPO component has a shape consistent with those measured in Intervals \#2-4 (Section~\ref{sec:Energy-dependent power and cross spectra}).

\textit{(4) Class-4, MJD 58305.6774$-$58305.85:}
In Interval~\#6 (Fig.~\ref{fig:pds_cs}, lower-right panel), taken in the SIMS, a Type-B QPO at $4.24\pm0.02$ Hz dominates the PDS and CS. The phase lag of the Type-B QPO is $-0.95_{-0.09}^{+0.05}$ rad and no Type-C QPO is detected in this observation, with a $\sim$95\% upper limit of the rms amplitude in the 0.5$-$12 keV band of 0.3\%.
We note that the phase lags of both the additional QPO component in the HIMS and the Type-B QPO in the SIMS, are consistent within 1$\sigma$ uncertainties. 

\subsection{Broadband rms vs. QPO frequency relation}
\label{sec:Type-B QPO in the HIMS}

\begin{figure}
	\includegraphics[width=\columnwidth]{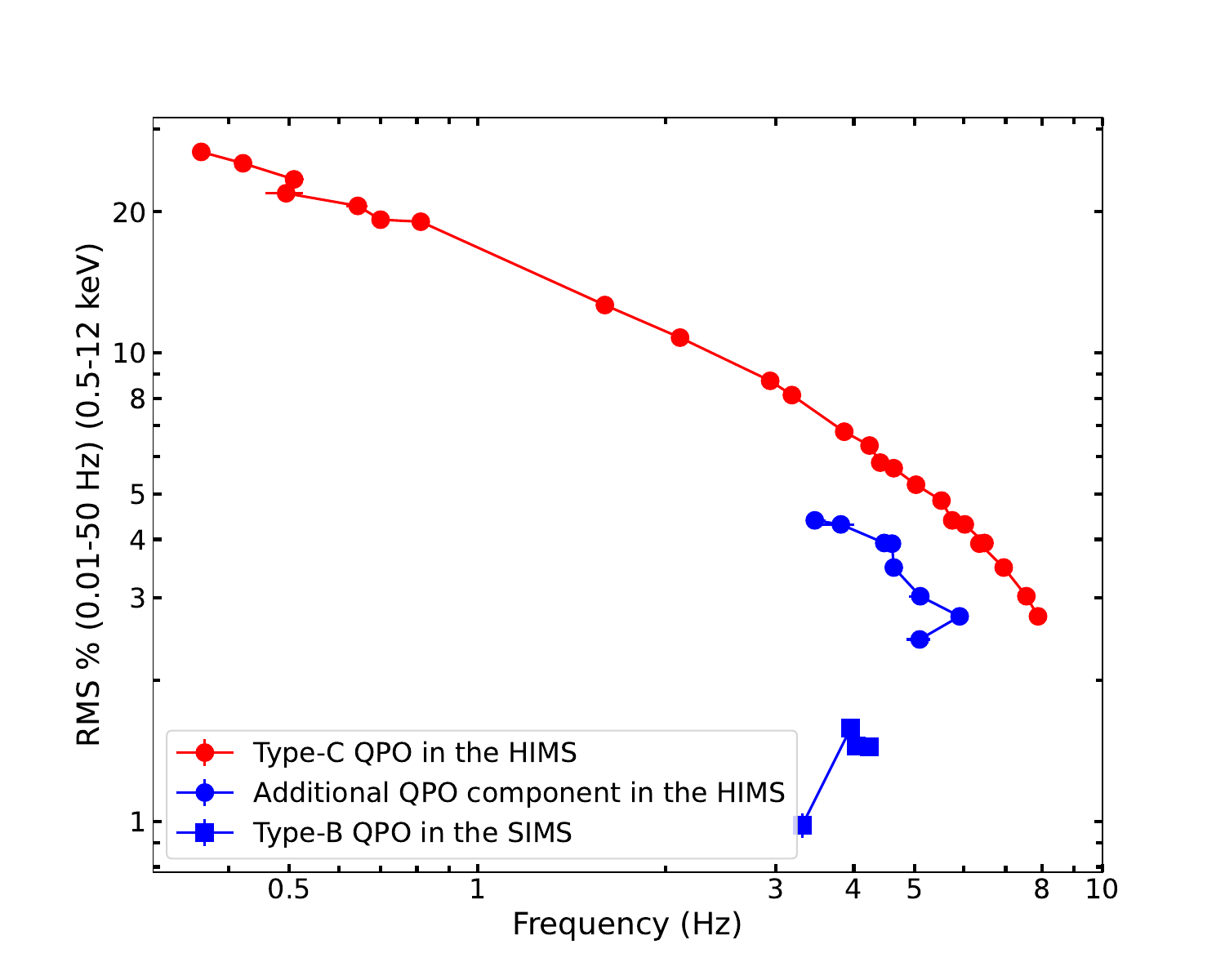}
    \caption{Broadband rms vs. QPO frequency relation for the QPOs in MAXI~J1820$+$070. The red circles are the Type-C QPO in the HIMS, the blue circles are the additional QPO component in the HIMS, and the blue squares are the Type-B QPO in the SIMS.} 
    \label{fig:rms_frequency}
\end{figure}

In Fig.~\ref{fig:rms_frequency}, we show the relation between the QPO frequencies and the broadband fractional rms in the 0.01$-$50 Hz range in the 0.5$-$12 keV energy band. For the Type-C QPO (red circles) observed in the HIMS, as the QPO frequency increases from $\sim$0.4 Hz to $\sim$8.0 Hz, the broadband fractional rms decreases from $\sim$30\% to $\sim$3\%. 
For the additional QPO component in the HIMS (blue circles), the QPO frequency varies between $\sim$3.5 Hz and $\sim$5.9 Hz, as the broadband fractional rms decreases from 4.5\% to 2.3\%.
For the Type-B QPO in the SIMS (blue squares), the QPO frequency varies between $\sim$3.3 Hz and $\sim$4.2 Hz, as the broadband fractional rms decreases from $\sim$1.6\% to $\sim$0.9\%.

\citet{2011MNRAS.418.2292M} first reported separate correlations between the QPO frequencies and the broadband fractional rms for different types of QPOs in GX~339$-$4. In that source, the Type-B QPO, detected only in the SIMS, when the PDS is characterized by a weak broadband noise, appears below the Type-C QPO branch. Similarly, as shown in Fig.~\ref{fig:rms_frequency}, the Type-B QPO observed in the SIMS of MAXI~J1820$+$070 also lies below the Type-C QPO branch.
\citet{2012MNRAS.427..595M} further reported the rms–frequency relation in GRO~J1655$-$40, including data from the ultra-luminous state (ULS), where the Type-C and Type-B QPOs appear simultaneously. In their case, the Type-B QPO is most often observed as a peaked-noise component in the PDS and has a lower frequency than the Type-C QPO, as in the HIMS of MAXI~J1820$+$070. Similarly, when the additional QPO component appears simultaneously with the Type-C QPO in the HIMS of MAXI~J1820$+$070, it occupies the same position in the broadband rms vs. QPO frequency relation as the Type-B QPOs in the ULS of GRO~J1655$-$40 \citep[see Figure 5 of][]{2012MNRAS.427..595M}.
Together with the comparison between the properties of this component and those of the Type-B QPO in the SIMS (see Sections~\ref{sec:Dynamical power spectrum} and~\ref{sec:Power and cross spectra}), we suggest that the additional QPO component may be related to the Type-B QPO observed in the SIMS, although the identification of the HIMS feature with the Type-B QPO remains subject to interpretation.

\subsection{Energy-dependent rms and phase-lag spectra of the QPOs}
\label{sec:Energy-dependent power and cross spectra}

\begin{figure}
	\includegraphics[width=1.05\columnwidth]{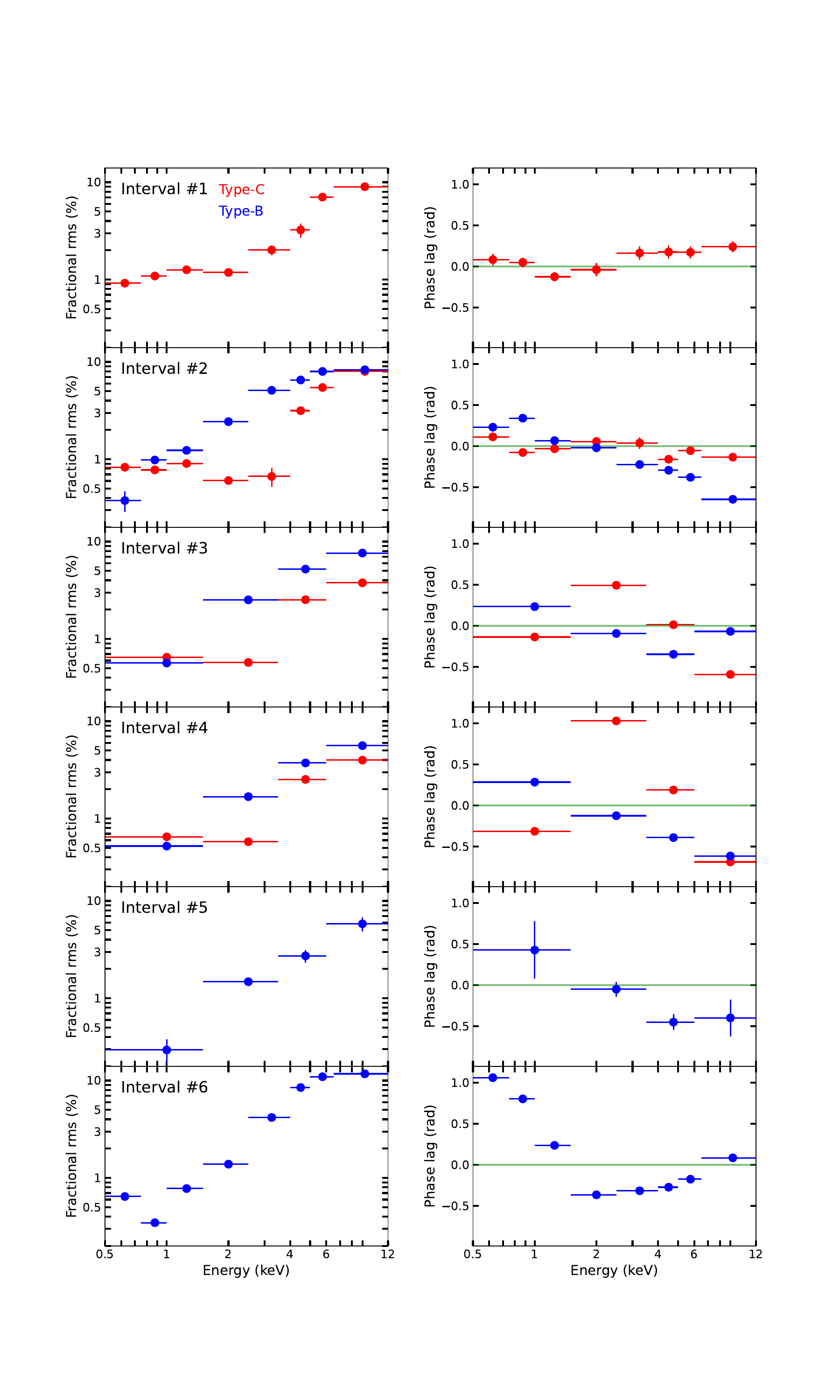}
    \caption{Fractional-rms (left panels) and phase-lag (right panels) spectra of the Type-C QPO (red) and the Type-B QPO (blue) of MAXI~J1820$+$070 in Intervals~\#1-6 in Fig.~\ref{fig:light_curve}, with 1$\sigma$ errors.} 
    \label{fig:rms_lag}
\end{figure}

In Fig.~\ref{fig:rms_lag}, we present the fractional-rms (left panels) and phase-lag (right panels) spectra of the Type-C QPO (red) in Intervals~\#1-4, the additional QPO component (blue) in the HIMS in Intervals~\#2-5, and the Type-B QPO (blue) in the SIMS in Interval~\#6.
For the Type-C QPO, the fractional-rms spectrum shows a similar trend across all observations: in Intervals~\#1-4, when the Type-C QPO is present, its fractional rms amplitude remains roughly constant between 0.5 and 4.0~keV and then increases above $\sim$4.0~keV, reaching a maximum near $\sim$10~keV. In Intervals~\#1 and \#2, the phase-lag spectrum of the Type-C QPO is more or less consistent with zero within 1$\sigma$ uncertainties. However, in Interval~\#3 and \#4, the phase-lag spectrum exhibits a peak around 2.0$-$3.0~keV, with a maximum value of $\sim$0.75~rad.
For the additional QPO component in the HIMS in Interval~\#2-5, the fractional rms amplitude increases steadily with energy and reaches a maximum at around $\sim$8~keV. Consequently, the fractional rms amplitude of the additional QPO component in the HIMS is significantly higher than that of the Type-C QPO in the $\sim$1.5$-$4.0~keV range, making the additional QPO component in the HIMS dominant in the 2$-$12~keV PDS when both QPOs appear simultaneously, as shown in Fig.~\ref{fig:pds_cs}. In Interval~\#2 and \#4, the phase lag of the additional QPO component decreases steadily across the 0.5$-$12.0~keV range, while in Interval~\#3 and \#5, the phase lag of the additional QPO component shows a minimum near $\sim$5~keV. 
The phase-lag spectrum of the Type-B QPO in the SIMS in Interval~\#6 also exhibits a minimum at around $\sim$2.5~keV, forming a clear ``U''-shaped trend~\citep[see also][]{2023MNRAS.525..854M} similar to those observed in other sources~\citep[e.g.;][]{2020MNRAS.496.4366B, 2021MNRAS.501.3173G, 2023MNRAS.519.1336P, 2023MNRAS.520.5144Z, 2026A&A...706A.208J}.

\section{Discussion}
\label{sec:discussion}

We present a timing analysis of the NICER observations in MAXI~J1820$+$070 during the HIMS-to-SIMS transition taking into account the power and cross spectra as in \citet{2024MNRAS.527.9405M} and \citet{2026A&A...706A.208J}. The transition is characterized by a sharp flux increase (Fig.~\ref{fig:light_curve}), and a rapid decrease in the strength of the broadband noise (Fig.~\ref{fig:dynamical_pds}). 
Approximately half a day before the transition, while the source is still in the HIMS and the PDS shows strong broadband variability, the Type-C QPO and its second harmonic,
we detect an additional QPO component with a characteristic frequency of $\sim$3.5$-$5.9 Hz in the high-energy PDS and the CS of MAXI J1820+070: 
(i) This additional QPO component appears to evolve continuously into the QPO identified as Type-B immediately after the transition (Figs.~\ref{fig:dynamical_pds} and~\ref{fig:pds_cs});
(ii) the comparison between the PDS in the interval immediately before the transition (Interval \#5) and that immediately after (Interval \#6) supports a continuous evolution between these features;
(iii) the location of this additional QPO component in the broadband rms vs. QPO frequency relation is consistent with the position of the Type-B QPOs in GRO~J1655$-$40~\citep{2012MNRAS.427..595M} and GX~339$-$4~\citep{2011MNRAS.418.2292M};
(iv) this additional QPO component is not harmonically related to the Type-C QPO (see Section~\ref{Peak feature in the hard-band PDS in the HIMS}).
All these properties indicate that this component is not a (sub)harmonic of the Type-C QPO, and suggest a possible connection with the Type-B QPO observed after the transition. Independently of its exact classification, the detection of this additional QPO component as a distinct feature, coexisting with and not harmonically related to the Type-C QPO, provides new constraints on the variability phenomenology during the HIMS-SIMS transition.
This identification would imply that, as it was previously observed in Swift~J1727.8$-$1613~\citep{2026A&A...706A.208J}, in MAXI~J1820$+$070 the emergence of the Type-B QPO may precede the transition from the HIMS to the SIMS, and therefore the Type-B QPO may be not connected to the discrete radio ejections, since in MAXI~J1820$+$070 those ejections appear at the HIMS-to-SIMS transition. On the contrary, the discrete radio ejections appear to be linked to the disappearance of the component of the accretion flow that, in the HIMS, produces the strong broadband variability and the Type-C QPO.

\subsection{The additional QPO component in the hard-band PDS in the HIMS}
\label{Peak feature in the hard-band PDS in the HIMS}

Besides the Type-C QPO and its second harmonic, we detect an additional QPO component in the high-energy PDS of MAXI~J1820$+$070 (Figs.~\ref{fig:dynamical_pds} and \ref{fig:pds_cs}), with a characteristic frequency of $\sim$3.5$-$5.9 Hz, while the source is still in the HIMS, approximately half a day before the transition to the SIMS.
The several properties of this additional QPO component are consistent with those of the Type-B QPO in the SIMS.
First, the centroid frequency of this additional QPO component ($\sim$3.5$-$5.9 Hz) is consistent with that of the Type-B QPO ($\sim$3.3$-$4.2 Hz) observed in the SIMS, suggesting that it could be the same feature that evolves smoothly during the transition (Figs.~\ref{fig:dynamical_pds} and \ref{fig:pds_cs}).
Second, the 2.0$-$12.0 keV vs. 0.3$-$1.5 keV broadband phase lag and the phase-lag spectrum of this additional QPO component are consistent with those of the Type-B QPO in the SIMS (Table~\ref{tab:rep_obs} and Fig.~\ref{fig:rms_lag}).
Third, a very similar case has been reported in GRO~J1655$-$40, although in the ULS rather than the HIMS, where the Type-C and Type-B QPOs appear simultaneously; as in MAXI~J1820$+$070, in that case, the Type-B QPO also appears as a narrow peak in the PDS and has a lower frequency than the simultaneous Type-C QPO. Moreover, the QPOs in GRO~J1655$-$40 and MAXI~J1820$+$070 occupy the same region in the plot of the broadband rms vs. QPO frequency.
Taken together, these results suggest that the additional QPO component observed in the HIMS of MAXI J1820$+$070 may be related to, or even be the precursor of, the Type-B QPO in the SIMS. However, alternative interpretations cannot be excluded.
Regardless of its exact classification, the detection of this additional QPO component as a distinct feature, coexisting with, and not harmonically related to, the Type-C QPO, provides new constraints on the variability phenomenology during the HIMS-SIMS transition.

Alternatively, this additional QPO component in the HIMS could be a sub-harmonic of the Type-C QPO.
However, several arguments disfavor this interpretation.
On the one hand, the averaged ratio between the frequencies of the additional QPO component and the Type-C QPO is significantly different from 0.5 (Fig.~\ref{fig:frequency_ratio}).
On the other hand, unlike the second harmonic, which always appears simultaneously with the Type-C QPO, we do not detect this additional QPO component in Interval~\#1, when the Type-C QPO and its second harmonic are strong in all energy bands.
One possibility is that this additional QPO component is already present at that stage but remains too weak to be detected, unlike the much stronger Type-C QPO and second harmonic. 
It is then unclear why the additional QPO component becomes relatively stronger, and even significantly stronger than the Type-C QPO in the high-energy PDS, as the Type-C QPO and its second harmonic weaken.
It is more likely that, as the Type-C QPO and its second harmonic weaken, either a previously hidden component becomes visible or a new component emerges.
In Interval \#5, just before the transition (lower-left panel of Fig.~\ref{fig:pds_cs}), the Type-C QPO and its second harmonic disappear, whereas the additional QPO component is significant in the hard-energy PDS.
In Swift~J1727.8$-$1613, \citet{2026A&A...706A.208J} found a Type-B QPO in the HIMS, with the frequency of the Type-B QPO evolving together with that of the Type-C QPO.
When Swift~J1727.8$-$1613 transitions briefly to the SIMS, the Type-C QPO disappears whereas the Type-B QPO remains.

All the above suggest that the additional QPO component in the HIMS of MAXI~J1820$+$070 is unlikely a sub-harmonic of the Type-C QPO.
We therefore put forward the possibility that, as in Swift J1727.8$-$1613, this additional QPO component in the HIMS of MAXI~J1820$+$070 may be, or may be related to, the Type-B QPO that appears in the SIMS.
We cannot, however, exclude the possibility that this feature is neither a Type-B nor a sub-harmonic of the Type-C QPO, but instead represents a totally different variability component that becomes prominent during the HIMS-SIMS transition. In this case, its similarity to Type-B QPOs may reflect similar physical conditions rather than a direct identification.

\subsection{Accretion-flow activity during the HIMS-to-SIMS transition}
\label{sec:What happen when a source transitions from the HIMS to the SIMS}

\citet[][]{2020ApJ...891L..29H} identified the HIMS-to-SIMS transition in MAXI~J1820$+$070 considering the sudden and simultaneous changes of the properties of the source in the radio and X-ray bands.
In the radio, discrete radio ejections were detected~\citep{2020ApJ...891L..29H, 2020NatAs...4..697B, 2021MNRAS.505.3393W}. In the X-ray band, the transition was characterized by a drop in the strength of the broadband noise and a switch from a Type-C to a Type-B QPO~\citep{2020ApJ...891L..29H, 2023MNRAS.525..854M}. Furthermore, the high-frequency covariance spectrum~\citep{2021A&A...654A..14D} suggested that, near the transition, the disk inner radius moved to, or very close to, the innermost stable circular orbit~\citep[see also,][]{2025MNRAS.538.1143L}.
However, our results show that an additional QPO component, with properties similar to those of the Type-B QPO, is present about half a day before the transition, and that at the transition itself it is the Type-C QPO that disappears at the same time that we observe a drop in the broadband noise fractional rms amplitude. 
This behavior is consistent with the findings in Swift~J1727.8$-$1613~\citep{2026A&A...706A.208J}, where the transition from the HIMS to a brief SIMS is marked by the disappearance of the Type-C QPO and a reduction in broadband-noise strength, whereas the Type-B QPO was already present during the HIMS.

The simultaneous increase in the Type-C QPO frequency and the decrease in the disk inner radius observed in MAXI~J1820$+$070 \citep{2025MNRAS.538.1143L} suggests a direct link between the Type-C QPO frequency and the truncation radius of the accretion disk, consistent with the predictions of the Lense–Thirring precession model~\citep{2009MNRAS.397L.101I}. 
In this scenario, the disappearance of the Type-C QPO accompanied by a marked drop in the broadband noise at the transition from the HIMS to the SIMS, when the disk reaches its minimum radius, would indicate that the hot inner flow responsible for the Type-C QPO contracts throughout the LHS and HIMS and vanishes at the transition to the SIMS. 
In the LHS and HIMS, a steady compact jet is observed~\citep[e.g.;][]{2021MNRAS.504.3862T, 2024ApJ...971L...9W}, whereas near the transition from the HIMS to the SIMS this steady jet is quenched and discrete radio ejections are observed~\citep{2020NatAs...4..697B, 2025ApJ...984L..53W}. 
We suggest that the simultaneous disappearance of the Type-C QPO and the steady jet could indicate that the poloidal magnetic field powering the compact jet is rooted in the hot inner flow responsible for the Type-C QPO. A similar idea was proposed by \citet{2006A&A...447..813F} and \citet{2020A&A...640A..18M}, who argued that a jet-emitting disk located within the standard accretion disk produces the steady jet and accounts for the Type-C QPO.
Our results suggest that the Type-C QPO, the accompanying strong broadband noise, and the steady jet are all produced in the same region located within the standard accretion disk, which collapses as the source transitions to the SIMS.

As discussed in this paper and in \citet{2026A&A...706A.208J}, the Type-B QPO, which could precede the transition from the HIMS to the SIMS, may not be directly associated with the discrete radio ejections, since these are launched during the transition. Instead, the radio ejections appear to be linked to the disappearance of the hot inner flow that, in the HIMS, produces the strong broadband variability and the Type-C QPO. As suggested by \citet{2026A&A...706A.208J} in the case of Swift~J1727.8$-$1613, the simultaneous radio ejection and soft X-ray flare observed at the transition may represent a channel for rapid energy release as the inner flow, thought to underpin the steady jet, the strong broadband variability, and the Type-C QPO in the HIMS, collapses.

During the process of the contraction and collapse of the hot inner flow accounting for the strong broadband variability and the Type-C QPO in the HIMS, as accreting material partly continues to fall toward the black hole, magnetic flux is advected inward and accumulates near the event horizon. The compressed and intensified magnetic field could then activate the Blandford–Znajek mechanism~\citep{2011MNRAS.418L..79T}, launching a new, relativistic jet located within the steady jet that is rooted in the hot inner flow responsible for the strong broadband variability and the Type-C QPO in the HIMS. This could lead to a spine–sheath jet, as the one observed in M87~\citep[][]{2018A&A...616A.188K}.
A similar two-component corona was proposed by \citet{2026A&A...706A.208J} to quantitatively explain the simultaneous presence of Type-C and Type-B QPOs in the HIMS of Swift~J1727.8$-$1613. In their picture, a disk–corona (the sheath) component that interacts directly with the accretion disk produces the Type-C QPO, while a jet-base corona (the putative spine) located within the disk–corona produces the Type-B QPO.

Our results raise questions about the traditional paradigm that associates Type-B QPOs exclusively with the SIMS and with discrete radio ejections. In MAXI J1820$+$070, we detect a QPO component during the HIMS with properties similar to those of the Type-B QPO, about half a day before the HIMS–to–SIMS transition, while the Type-C QPO and strong broadband noise are still present. This timing rules out interpretations in which the Type-B QPO is produced by the discrete radio jet itself, for example through precession of the jet~\citep{2016MNRAS.460.2796S, 2020A&A...640L..16K}, since the discrete radio ejections occur only at the transition. Instead, the radio ejections appear to be linked to the disappearance of the component of the accretion flow that, in the HIMS, produces the strong broadband variability and the Type-C QPO. The fact that the Type-B-like QPO (the additional QPO component) is not detected in the soft band during the HIMS suggests that the region where this QPO is produced does not interact directly with the accretion disk at that stage and lies closer to the black hole than the region producing the Type-C QPO. In the same spirit as proposed for Swift J1727.8$-$1613, this is consistent with the idea that the Type-B QPO traces a distinct, compact region of the inner flow, often discussed as a “jet-base corona”, that can coexist with the larger-scale disk–corona responsible for the Type-C QPO. In that context, it is tempting to speculate that a two-component outflow, e.g. a spine–sheath-like configuration~\citep[][]{2018A&A...616A.188K}, could develop during the contraction and collapse of the inner flow, with the Type-B QPO associated with the more compact, inner region. Our data provide no direct evidence for a spine–sheath structure in MAXI J1820$+$070, and the point here is only that the simultaneous presence of Type-C and Type-B-like QPOs in the HIMS may be qualitatively compatible with such a picture. After the transition, when the strong broadband noise and the Type-C QPO have disappeared, the Type-B QPO remains the only prominent timing feature in the SIMS.

\section{Conclusions}

Using NICER observations, besides the Type-C QPO and its second harmonic, we detect an additional QPO feature at $\sim3.5-5.9$ Hz in the high-energy PDS of MAXI J1820$+$070 in the HIMS that is not harmonically related to the Type-C QPO but shares several properties with the Type-B QPO observed in the SIMS. Our results suggest that this additional QPO could be related to, or even be a precursor of, the Type-B QPO that is observed later, when the source is in the SIMS. Alternative interpretations, however, cannot be excluded.
Together with a similar case reported in Swift~J1727.8$-$1613, this finding suggests that in BHXBs the Type-B QPO may emerge prior to the HIMS-to-SIMS transition, and would therefore not be confined exclusively to the SIMS.
These findings imply that the bright discrete radio ejections detected during the HIMS-to-SIMS transition could be linked to the collapse of the so-called disk corona, which produces the Type-C QPO and the strong broadband noise in the LHS and HIMS, rather than to the emergence of the Type-B QPO.
We speculate that the simultaneous presence of Type-C and Type-B QPOs in the HIMS could be compatible with a spine-sheath outflow structure, in which the spine produces the precursor of the Type-B QPO in the HIMS, corresponding to the so-called jet-base corona, while the sheath produces the Type-C QPO, the accompanying strong broadband noise, and the steady jet, corresponding to the so-called disk corona.

\begin{acknowledgements}
We thank the referee for the insightful suggestions that helped improve the clarity of our work. MM acknowledges the research programme Athena with project number 184.034.002, which is (partly) financed by the Dutch Research Council (NWO). PJ acknowledges support from the China Scholarship Council (CSC 202304910058). FG acknowledges support by PIBAA 1275 and PIP 0113 (CONICET). FG was also supported by grant PID2022-136828NB-C42 funded by the Spanish MCIN/AEI/ 10.13039/501100011033 and “ERDF A way of making Europe”. RM acknowledges support from the Royal Society Newton Funds.
\end{acknowledgements}

\bibliographystyle{jwaabib}
\bibliography{references}

\appendix

\section{PDS and CS in Intervals \#3 and \#4}
\label{sec:PDS and CS in Obs 3 and 4}

\begin{figure}
	\includegraphics[width=0.9\columnwidth]{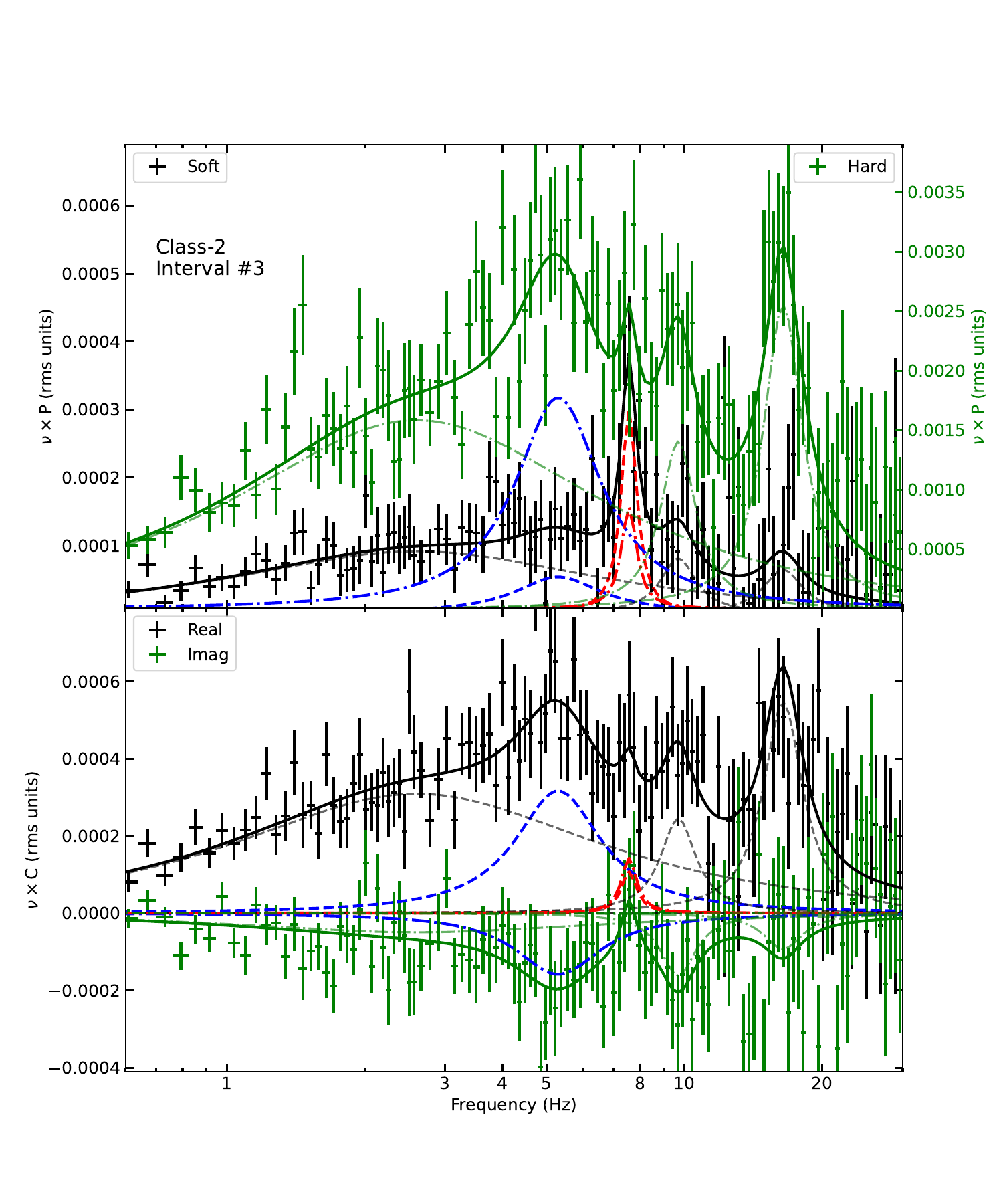}
    \includegraphics[width=0.9\columnwidth]{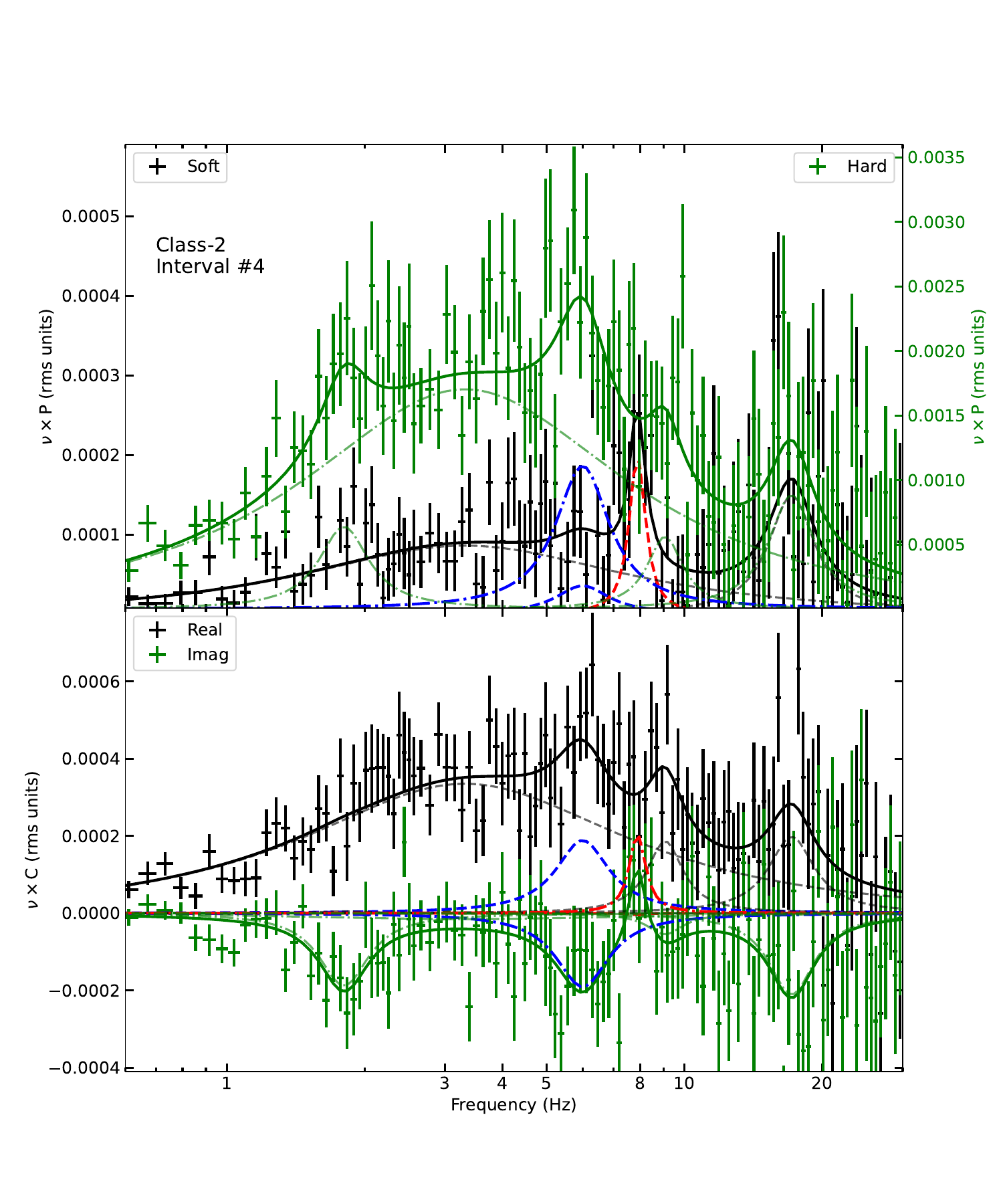}
    \caption{Same as Fig.~\ref{fig:pds_cs}, but for Intervals~\#3 (upper panel) and~\#4 (lower panel) shown in Fig.~\ref{fig:light_curve} and classified as Class-2 PDS.} 
    \label{fig:pds_cs_obs3_4}
\end{figure}

In Fig.~\ref{fig:pds_cs_obs3_4}, we present the fit results from Intervals~\#3 (upper panel) and \#4 (lower panel), classified as Class-2.
Interval~\#3 shows a Type-C QPO at $7.55_{-0.05}^{+0.09}$~Hz with a phase lag of $0.9_{-0.5}^{+0.4}$~rad. Its second harmonic peaks at $16.3_{-0.1}^{+0.2}$~Hz with a phase lag of $-0.17_{-0.14}^{+0.07}$~rad, and another QPO is seen at $5.1 \pm 0.2$~Hz with a phase lag of $-0.46_{-0.08}^{+0.18}$~rad.
Interval~\#4 also shows a Type-C QPO at $7.9_{-0.2}^{+0.1}$~Hz with a phase lag of $1.6_{-0.7}^{+0.3}$~rad. The second harmonic peaks at $17.0 \pm 0.6$~Hz with a phase lag of $-0.8 \pm 0.2$~rad, and another QPO is detected at $5.9 \pm 0.1$~Hz with a phase lag of $-0.80_{-0.2}^{+0.1}$~rad.
Both Intervals~\#3 and \#4 exhibit PDS and real and imaginary parts of the CS similar to those in Interval~\#2, but with higher characteristic frequencies of all variability components.


\end{document}